\documentclass[twocolumn,english,aps,prl,groupedaddress,superscriptaddress,longbibliography]{revtex4-1}
\usepackage{graphicx,epsfig,units}
\usepackage{xcolor} 
\usepackage{soul} 
\usepackage{amsmath,amsfonts,mathrsfs,amsbsy,bm,babel}
\usepackage{changes}
\usepackage{natbib}

\begin{document}
\title{Metamagnetism of few layer topological antiferromagnets}

\author{C. Lei}
\affiliation{Department of Physics, The University of Texas at Austin, Austin, TX 78712}

\author{O. Heinonen}
\affiliation{Argonne National Laboratory, Argonne, IL USA}

\author{A.H. MacDonald}
\affiliation{Department of Physics, The University of Texas at Austin, Austin, TX 78712}

\author{R.~J.~McQueeney}
\affiliation{Ames Laboratory, Ames, IA, 50011, USA}
\affiliation{Department of Physics and Astronomy, Iowa State University, Ames, IA, 50011, USA}

\date{\today}

\begin{abstract}
MnBi$_2$Te$_4$ (MBT) is a promising antiferromagnetic topological insulator whose films provide access to novel and technologically important 
topological phases, including quantum anomalous Hall states and axion insulators.
MBT device behavior is expected to be sensitive to the 
various collinear and non-collinear magnetic phases that are accessible in applied magnetic fields. Here, we use classical Monte Carlo simulations and electronic structure models to 
calculate the ground state magnetic phase diagram as well as topological and optical properties for few layer films with thicknesses
up to six septuple layers.  Using magnetic interaction parameters appropriate for MBT, 
we find that it is possible to prepare a variety of different magnetic stacking sequences,
some of which have sufficient symmetry to disallow non-reciprocal optical response and 
Hall transport coefficients.  Other stacking arrangements do yield large Faraday
and Kerr signals, even when the ground state Chern number vanishes.
\end{abstract}
\maketitle

\section{Introduction}
MnBi$_2$Te$_4$ (MBT) is a promising platform for the development of unique devices based on topological electronic bands \cite{Eremeev2017,Otrokov_2017,Otrokov19,Rienks2019,Chen2019,Hao2019,Li2019,Swatek2020,Lei20,Lee19,Zhang20,Gong19,Zhang2019,Klimovskikh2019,Li2019_theory,Chowdhury_2019,INS,Yan19,Zeugner2019,Wu2019,Zhang2019_AHC,Otrokov2019_film,Liu20,Chen2019_Pressure,Ding2020,Vidal2019,Vidal2019b,Ge20}.  The utility of MBT is consequence of its natural layered structure, which consists of stacks of ferromagnetic (FM) septuple layers (SLs) with out-of-plane magnetization and with inverted electronic bands with non-trivial topology 
\cite{Eremeev2017,Otrokov_2017,Otrokov19}.  There is great interest in manipulating the sequence of magnetic and topological layers as a means to control phenomena related to the band topology \cite{Lei20}. Bulk MBT adopts a staggered antiferromagnetic (AF) stacking of the FM SLs \cite{Otrokov19, Yan19, Lee19, Gong19}, providing the first realization of an AF topological insulator, which is predicted to host unique axion electrodynamics\cite{Essin09, Mong10}.   Weak interlayer magnetic interactions across the van der Waals gap and uniaxial magnetic anisotropy allow for facile control of the magnetic structure. For example, a bulk Weyl semimetallic phase is predicted when all FM layers in MBT are co-aligned with a small applied magnetic field \cite{Li2019_theory,Chowdhury_2019,Lei20}.

Perhaps the most exciting opportunity in MBT materials is the 
possibility of developing thin film devices with precisely controlled magnetic stacking sequences. 
In this context, devices with even and odd numbers of AF stacked layers offer 
qualitatively different topological phase possibilities
\cite{Zhang20, Liu20, Ge20, Deng20, Ovchinnikov20,Yang20}.   
An odd number of magnetic layers necessarily has partially compensated magnetization 
that is beneficial for the observation of the quantum anomalous Hall (QAH) effect \cite{Zhang20}. 
On the other hand, fully compensated magnetization occurs in even layer devices in the absence of an applied magnetic field
and provides an ideal platform to search for axion insulators with 
quantized magnetoelectric coupling \cite{Liu20,Zhang2019}. However, there are also much richer possibilities in both
odd or even layer samples \cite{Yang20}, since the application of a magnetic field can result in different collinear magnetic stackings with partially compensated magnetization, or even to non-collinear (canted) magnetic phases.   Here, we consider the possibility to stabilize such phases and the potential for realizing unique topological phases in this scenario.  

The magnetization behavior of AF magnetic multilayers displays rich metamagnetic behavior with features, such as surface spin-flop transitions, that are not observed in bulk AFs \cite{Hellwig03, Bogdanov}.  This complex behavior is a consequence of competition between single-ion anisotropy, interlayer magnetic exchange, and Zeeman energy.  For topological materials, whether the symmetry of different magnetic layer stacking sequences can affect the topological properties of the bands is an open question.  For example, collinear metamagnetic phases with the same net magnetization can result from magnetic layer stackings that may or may not break mirror symmetry ($\mathcal{M}$).  
Here we use classical Monte Carlo simulations to show that it is generally possible to tune-in 
different stacking sequences with distinct symmetries.  What's more,
we find that realistic values of the single-ion anisotropy and 
exchange place MBT close to this tunability regime.  
Finally, we discuss the topological properties of accessible field-tuned states and strategies to identify them experimentally.

\section{Monte Carlo Simulations of Bulk MnBi$_2$Te$_4$}

\begin{figure}
\includegraphics[width=1.0\linewidth]{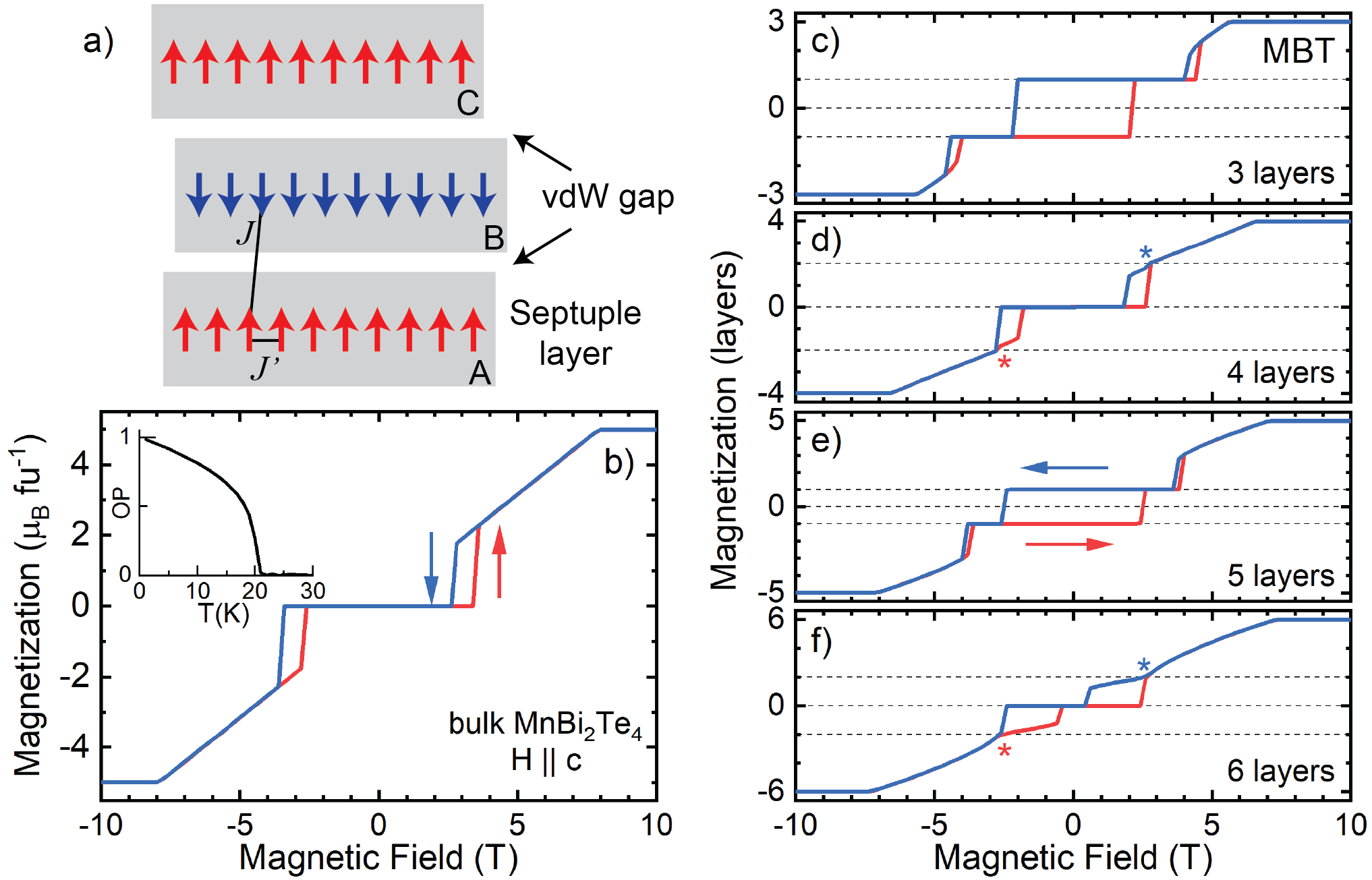}
\caption{\footnotesize (a) Schematic magnetic layer structure of MBT in the crystallographic unit cell with close-packed stacking of septuple layers.  The intralayer ($J'$) and interlayer ($J$) magnetic interactions are indicated.  The gray shaded boxes indicate the full septuple layers separated by a van der Waals gap. (b) Monte Carlo simulations of the bulk magnetization of MBT with field applied perpendicular to the Mn layers.  Inset shows the order parameter (OP) of the staggered A-type AF order as a function of temperature. (c)-(f) Monte Carlo simulations of the magnetization of 3, 4, 5, and 6 layer MBT, respectively, using the nominal Heisenberg parameters.  For $N=4$ and 6, the "$*$" indicates an additional phase transition.}
\label{fig:magnetization}
\end{figure}

The spin lattice of MBT consists of triangular FM layers which are stacked in a close-packed fashion along the direction perpendicular to the layers.  Interlayer interactions are AF, resulting in the zero-field A-type ground state [see Fig. \ref{fig:magnetization}(a)].  Magnetization and neutron diffraction experiments find Mn moments oriented perpendicular to the layers consistent with uniaxial magnetic anisotropy \cite{Yan19}.  These features suggest that a simple spin model can be used to study 
the magnetization behavior of MBT in an applied magnetic field
\begin{equation}
H= J' \sum_{\langle ij \rangle||} \textbf{S}_{i} \cdot \textbf{S}_{j} + J \sum_{\langle ij \rangle\perp}\textbf{S}_{i} \cdot \textbf{S}_{j} - D \sum_{i} S_{i,z}^{2} - g\mu_{\rm B} \textbf{H} \cdot \sum_{i} \textbf{S}_{i}.
\label{heisenberg}
\end{equation}
Here, $\textbf{S}_{i}$ is the Mn spin at site $i$ ($S=5/2$), $J'<0$ is the intralayer FM exchange, 
$J>0$ is the interlayer AF exchange and $D$ is the uniaxial single-ion anisotropy.  Each Mn ion has six interlayer and intralayer nearest-neighbors ($z=6$). A representative and consistent set of magnetic coupling 
parameter values for MBT have been obtained from magnetization \cite{Yan19} and inelastic neutron scattering experiments \cite{INS}; $SJ'=-0.35$~meV, $SJ=0.088$~meV, and $SD=0.07$~meV.  

Using these nominal values, classical Monte Carlo (MC) simulations on bulk and few layer MBT systems  have been performed using both UppASD \cite{UppASD} and Vampire \cite{Vampire} software packages. 
MC simulations are first performed on bulk MBT with a $21 \times 21 \times 12$ 
system size (15876 spins) with periodic boundary conditions.  MC simulations are run with the field pointed perpendicular to the layers using 50000 MC steps per field.   To account for hysteresis and history dependence, we begin the simulations with the field polarized state at $H=$ 10 T and ramp the field down in equal steps, 
using the final state from the previous field as the initial state for the current field.  

Figure \ref{fig:magnetization}(b) summarizes typical 
MC simulation results for bulk MBT that reveal a N\'{e}el temperature of 22 K and magnetization 
curves with field-polarized saturation fields ($\mu_0 H_{ab}^{sat}=10.3$~T and 
$\mu_0 H_c^{sat}=7.9$~T), and a spin-flop field ($\mu_0 H_{SF}=3.7$~T), in agreement with bulk measurements \cite{Yan19, Lee19, Gong19}.  
The hysteresis of the spin-flop transition is noticeably absent in the experimental data of bulk single-crystals, which suggests that the spin-flop transition in real crystals does not occur by coherent layer rotation and could occur instead by surface nucleation \cite{Sass20}.

For a general bulk uniaxial AF, critical values of the ratio $D/zJ$ determine the magnetization behavior when the field is applied along the $c$-axis. Only three phases are possible, the AF phase, the canted spin-flop phase (SF), and the field-polarized phase (FM). For relatively weak anisotropy $D/zJ<1/3$, a first-order spin-flop transition phase is expected, followed by a second-order transition to the field polarized state (AF $\rightarrow$ SF $\rightarrow$ FM).   
The nominal MBT parameters yield $D/zJ \approx 0.13$
which is within the spin-flop regime.  For dominant single-ion anisotropy $D/zJ > 1$, the virgin AF state is swept out by a weak applied field and a FM hysteresis loop develops.  This behavior is observed in MnBi$_4$Te$_7$, in which the addition of a non-magnetic Bi$_2$Te$_3$ spacer between
MnBi$_2$Te$_4$ layers dramatically weakens the interlayer magnetic exchange \cite{Tan20, Wu20}.  
In the intermediate regime $1/3 < D/zJ < 1$, the AF $\rightarrow$ FM transition occurs directly (a spin-\textit{flip} transition), whereas the field-reversed transition goes through the spin-flop regime (FM $\rightarrow$ SF $\rightarrow$ AF). 

\section{Field-Tuned Few-Layer Magnetization}

We now consider the behavior of the magnetization in thin film 
samples consisting of $N$ magnetic layers, with $N=3,4,5$ and 6.  The magnetic phase diagram of the $N$-layer systems is much more complex than the bulk phase diagram because regions of stability exist (as described in detail below) that correspond to collinear magnetic phases with partially compensated magnetization (ferrimagnetic phases).  In this respect, significant differences occur between odd or even layer systems because the net magnetization cannot be fully compensated in odd layer films.  Fig. \ref{fig:magnetization} (c)-(f) shows the magnetization for $N=3$, 4, 5, and 6-layer MBT obtained from MC simulations using a
$11 \times 11$ system size for the basal layer and the nominal MBT Heisenberg parameters.  Select simulations with a $21 \times 21$ basal layer produced no significant changes in the magnetization sweeps. 

For $N=$ 3 and 5 layer simulations with the nominal MBT parameters, we find the expected FM hysteresis loop corresponding to the magnetization of a single uncompensated layer ($M=\pm1$). In addition, we find a hysteretic spin-flop-like transition from the $M=\pm1$ phase to the fully polarized FM phase.  For $N=4$ and $N=6$, magnetization curves resemble bulk MBT with a AF $\rightarrow$ SF $\rightarrow$ FM sequence of transitions.  However, one notices evidence for an additional transition [indicated with a "$*$" in Fig.~\ref{fig:magnetization}(d) and (f)] within the spin-flop phase near 3 T.  As we will show below, this transition demonstrates that the nominal parameters of MBT are close to a critical point in the phase diagram at which the $M=2$ collinear phase becomes stable.  We note that experimental evidence exists for $M=2$ magnetization plateaus in the $N=4$ and 6-layer films based on reflective magnetic circular dichroism experiments \cite{Yang20,Ovchinnikov20}.  The experimental magnetization results in Refs.~\cite{Yang20,Ovchinnikov20} are analyzed using numerical methods similar to the Mills model \cite{Mills68} described below.

\begin{figure}
\includegraphics[width=1.0\linewidth]{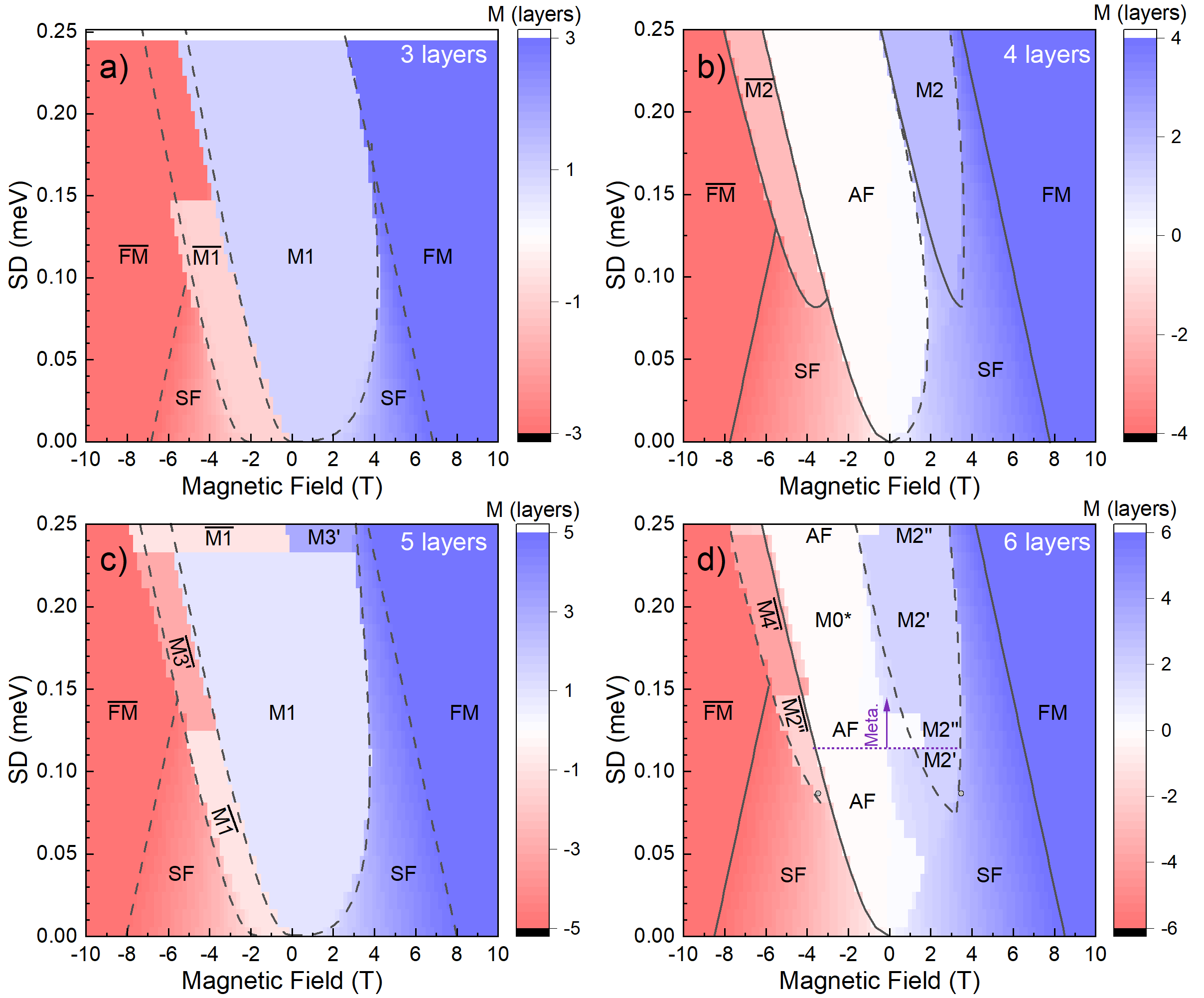}
\caption{
\footnotesize  Phase diagrams obtained by MC simulation showing the layer magnetization of a uniaxial layered antiferromagnet {\it vs.} magnetic field applied perpendicular to the layers and single-ion anisotropy strength
($SD$) for (a) 3 layers, (b) 4 layers, (c) 5 layers and (d) 6 layers.  Simulations start in the FM phase at 10 T and the field strength is reduced in equal steps of 0.2 T to -10 T. 
The solid lines are phase boundaries and metastability limits obtained from the Mills model,
the dashed lines are guides to the eye.  Collinear phases are labeled as described in the text.  
For $N=6$, panel (d) indicates a metastability limit below which only the M2$'$ phase is 
stabilized out of the SF phase.}
\label{fig:phase_diagram}
\end{figure}

To explore the nature of this critical point and parameter regimes beyond the nominal values 
chosen to represent MBT, we calculated the phase diagrams for the 3, 4, 5, and 6-layer systems 
as a function of field and the single-ion anisotropy parameter, as shown in Fig.~\ref{fig:phase_diagram}.  
Generally, these phase diagrams show regions of collinear magnetism separated by non-collinear (spin-flop-like) phases.  For the largest uniaxial anisotropy values, the system approaches the behavior of a finite Ising chain where successive first-order transitions occur via single-layer spin-flips. 
We label the collinear ground state phases as; AF ($M=0$, $N=even$ only), M$n$ ($M=n$, with $n-1$ broken AF bonds), and FM ($M=N$, field-polarized with all AF bonds broken). 
Their time-reversed states are indicated by a bar (eg. $\overline{{\rm M}2}$) and states 
that have identical ground state energies within our model, when they occur, 
are differentiated by a prime symbol (eg. M2$'$, M2$''$).  
Metastable excited states, when they occur, are labeled with an 
asterisk (eg. M0$^*$).

\section{Metamagnetic states}

\begin{figure}
\includegraphics[width=1.0\linewidth]{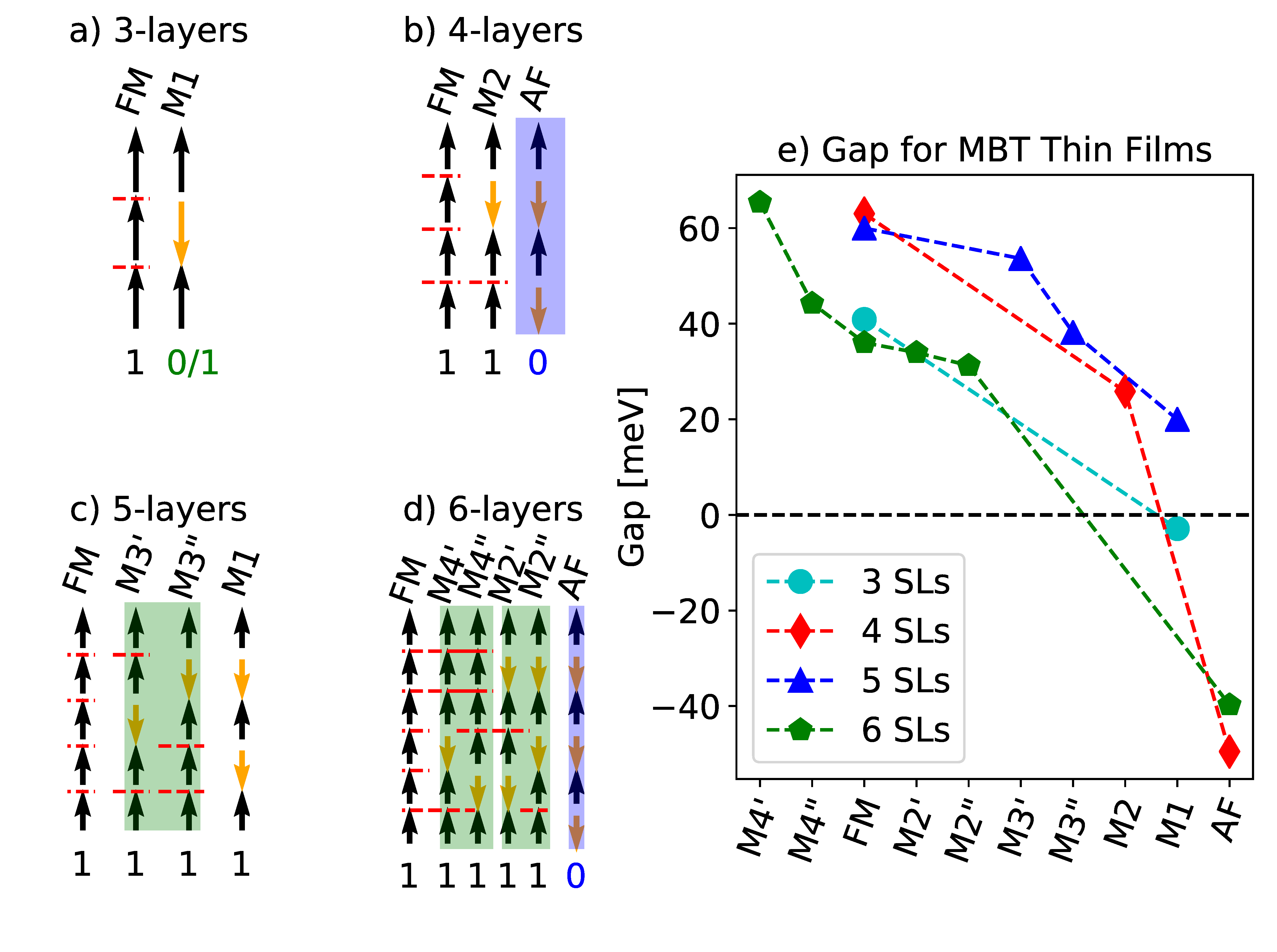}
\caption{\footnotesize Possible collinear magnetic ground states for $N=3$, 4, 5, and 6-layer systems, labeled as either FM, AF or M$n$, where $n$ is the uncompensated net layer magnetization.  Red dashed lines indicate broken AF bonds that cost an exchange energy of $3J$ each.  Shaded green rectangles enclose degenerate states with the same magnetization, but different stacking sequences indicated by the prime symbol. Shaded blue rectangles enclose the states with odd-parity magnetic configuration (i.e. $\mathcal{T}\mathcal{M}$ symmetrized magnetic configuration, with $\mathcal{T}$ time-reversal and $\mathcal{M}$ mirror symmetry).  The integers below each state denote their Chern numbers calculated from a the simplified Dirac-cone model and (e) are the calculated band gaps {\it vs.} the magnetic stacking sequences of few-layer MBT thin films.
}
\label{fig:mag_stacking}
\end{figure}

The distinct collinear ground states that occur for $N=3$ and $N=4$ are illustrated in Fig.~\ref{fig:mag_stacking}(a) and (b).
For $N=4$, the MC phase diagram in Fig.~\ref{fig:phase_diagram}(b) is consistent with the phase boundaries, critical points, and metastability limits obtained using the Mills model \cite{Mills68, Bogdanov}.  As expected, slightly increasing $D$ from the nominal MBT value of $SD=$ 0.07 meV stabilizes the M2 collinear phase, replacing the inflection in Fig.~\ref{fig:magnetization}(d) with a magnetization plateau.  Analysis of 4-layer Mills model reveals that this critical point occurs at $(SD,\mu_0 H)_{\beta}=$ (0.082 meV, 3.59 T) \cite{Bogdanov}, corresponding to a critical ratio of $D/zJ=0.16$.  

For odd thin film thicknesses ($N=3$ and 5), 
the magnetization sequence at low $D/zJ$ reveals M1 and $\overline{{\rm M}1}$ phases that form a hysteresis loop.  
For $N=5$, larger $D/J$ reveals two M3 states occur, labelled as M3$'$ and M3$''$  in Fig.~\ref{fig:mag_stacking}(c), 
which have identical energies in our model but are distinguished by the presence (M3$'$) or absence (M3$''$) of mirror symmetry.
Analysis of the sublattice magnetization from our simulations shows that only the mirror symmetric M3$'$ phase appears in the 
range of $D/zJ$ studied.

Much more interesting and complex behavior is obtained for thin films with $N=6$, where collinear phases appear 
that have equal uncompensated magnetizations but different symmetries.
These phases appear beyond the critical point, as in the $N=4$ case, which is 
estimated to be $(SD,\mu_0 H)_{\beta}=$ (0.084 meV, 3.46 T) from the 6-layer Mills model \cite{Bogdanov} with $D/zJ=0.16$.  
Close to the critical point, two different M2 phases appear at positive and negative fields, labeled M2$'$ and M2$''$ as shown in
Fig.~\ref{fig:mag_stacking}(d).  The mirror symmetric M2$'$ phase has a single broken AF bond in the center of the stack,
and is found emerging out of the spin-flop phase (FM$\rightarrow$SF$\rightarrow$M2$'$).  The mirror symmetry broken M2$''$ has a broken 
AF bond on the surface and is always stabilized out of the AF state (AF$\rightarrow$M2$''$). 

\begin{figure}
\includegraphics[width=1.0\linewidth]{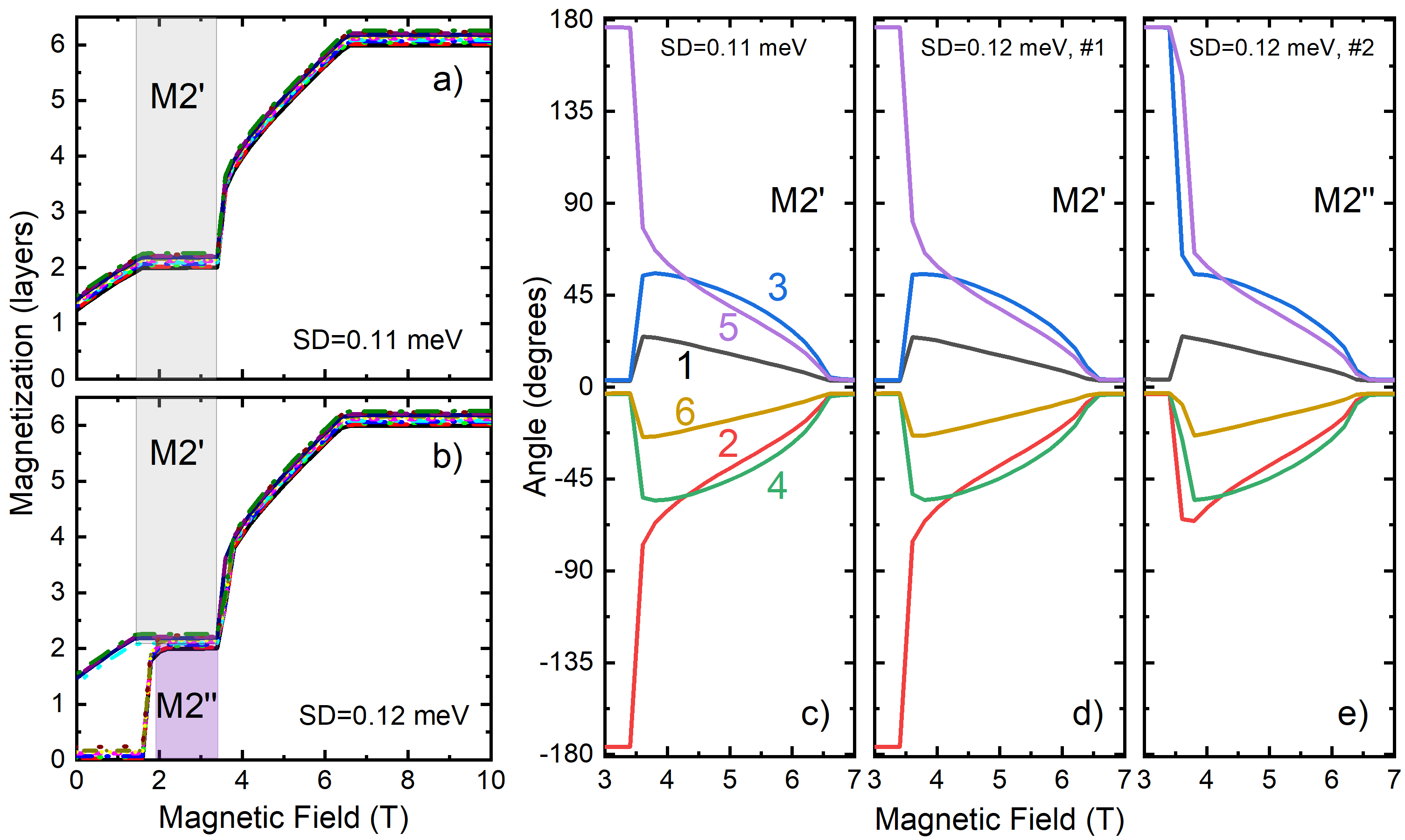}
\caption{\footnotesize (a)-(b) Twelve MC simulations for $N=6$ repeated under
identical starting conditions near the metastability limit for formation of the M2$'$ phase.  
Curves have a slight vertical offset for clarity. (a) For $SD=0.11$ meV, only the M2$'$ phase appears. 
(b) For $SD=0.12$ meV, either the M2$'$ or M2$''$ phase appears.  
(c)-(e) show the evolution of the magnetization angle for each layer
for different simulations.  (c) For $SD=$ 0.11 meV, which is below the metastability limit, 
the angle of layers 2 and 5 in the spin-flop phase approaches 90$^{\circ}$ before flipping into the M2$'$ phase.  
For $SD=$ 0.12 meV, different simulations (labeled \#1 and \#2) result in either 
(d) M2$'$ with layers 2 and 5 flipping or (e) M2$''$ with layers 3 and 5 flipping.}
\label{fig:metastability}
\end{figure}

The preference for the AF$\rightarrow$M2$''$ transition over the AF$\rightarrow$M2$'$ transition can be understood from metastability arguments related to the barrier height for layer flips that is determined primarily from the uniaxial anisotropy.  The AF$\rightarrow$M2$''$ transition requires a coherent spin flip of the surface layer only, whereas AF$\rightarrow$M2$'$ requires three layer flips.  From the FM side, both M2$'$ and M2$''$ require two layer flips and, for this reason, either phase is likely to appear within our simulations when $D/zJ \gtrsim 0.2$, as indicated by the purple dotted line in Fig.~\ref{fig:phase_diagram}(d).  Due to its larger barrier height, M2$'$ has a lower metastability field to enter the AF phase than M2$''$ as the field is reduced.  To illustrate the metastability limit, Fig.~4(a) and 4(b) show that repeated simulations in which the field is reduced starting
the FM phase will always generate M2$'$ with $SD=0.11$ meV, whereas either M2$'$ or M2$''$ may appear with $SD=0.12$ meV.  This metastability
limit at intermediate $D/zJ$ originates from the intervening spin-flop phase that selectively lowers the barrier to the mirror symmetric M2$'$ phase when layers 2 and 5 have a large spin-flop angle, as shown in Fig.~\ref{fig:metastability}(c)--(e).  

At large $D/zJ$, metastability issues with the MC simulations reveal that even excited states (such as the M0$^*$ state with stacking sequence up-down-up-down-down-up) may be trapped in a local minimum. The occurrence of the M0$^*$ phase is dependent on whether the M2$'$ or M2$''$ phase appears when lowering the field out of the FM phase. As described above, when the M2$''$ phase appears, the AF phase is favored since only a spin flip of the surface layer is required.  When the M2$'$ phase appears, transition to the AF phase has a high barrier requiring three layer spin flips.  It is therefore more likely for the M2$'$ phase to transition to the metastable M0$^*$ state 
in which the barrier height is set by a single layer flip.  While this regime is not applicable for MnBi$_2$Te$_4$, where $D/zJ \approx 0.13$, it may be applicable to MnBi$_4$Te$_7$ in which non-magnetic Bi$_2$Te$_3$ spacer layers dramatically reduce $J$.

\section{Topological and Optical Properties}

A very interesting question is how the stacking sequence of the magnetic layers in MBT thin films and the possible concomitant breaking of symmetries affect observable electronic properties, particular those related to 
the topological classification of the electronic structure. 
To gain insight, we used a simple model of stacked 2D Dirac metals, 2 for each MBT layer, to
calculate the band structure, Chern numbers, and magneto-optical responses 
of different magnetic states.
Previous work \cite{Lei20} has shown that with appropriate coupling 
between the Dirac metals, models of this type provide a reasonable description of MBT thin films.

As summarized in Fig. \ref{fig:mag_stacking}, we find that for $N >3$ all states have unity-magnitude Chern numbers 
whenever they have uncompensated magnetization, signaling non-trivial topological states.
The calculated Chern numbers are listed below each state. Odd $N >3 $ systems are therefore always QAH insulators.
The $N=3$ M1 state is labelled as $0/1$ because it has an extremely small gap and can be on either side
of the topological phase, transition depending on exchange interaction parameters and on 
the residual electric field that is likely to be present in any MBT thin films \cite{Lei21_QAH}. 
The gaps of the various magnetic states we have identified are shown in Fig. \ref{fig:mag_stacking}(e),
with a negative sign attached to distinguish cases in which the Chern number is $0$ from cases in which the Chern number is $1$.
The QAH gaps are generally larger for FM magnetic configurations for MBT thin films, 
except for the M4$'$ and M4$''$ states of $N=6$ films, which have larger gaps than the M6 state.

Symmetries play an important role in film electronic properties.  We define odd-parity magnetic configurations 
as ones that have the property that the magnetic moments reverse upon layer reversal (i.e.~$\mathcal{T}\mathcal{M}$ symmetrized magnetic configurations with $\mathcal{T}$ the time-reversal operator).  Odd-parity magnetic configurations 
can be read from the cartoons enclosed with shaded blue rectangles in Fig. \ref{fig:mag_stacking}, and include even--$N$ AF configurations.
Whenever the magnetic configuration has odd parity, the band Hamiltonian is invariant under the product of time-reversal symmetry $\mathcal{T}$ and inversion symmetry $\mathcal{I}$.  
(Note that the spin-orbit coupling terms in coupled Dirac-cone model on the top and bottom surfaces of each septuple layer differ by a sign.)
For many observables, the consequences of $\mathcal{T} \mathcal{I}$ invariance are the same as the consequences separate $\mathcal{T}$ and $\mathcal{I}$ invariance.  
For example, $\mathcal{T} \mathcal{I}$ invariance implies that the Berry 
curvature $\Omega_{n}(\mathbf{k}) = -\Omega_{n}(\mathbf{k})=0$.
The Berry curvature therefore vanishes 
identically and a generalized Kramer's theorem implies that all bands are 
doubly degenerate.  It follows that the Chern number vanishes for odd-parity magnetic configurations.

\begin{figure}
\includegraphics[width=1.0\linewidth]{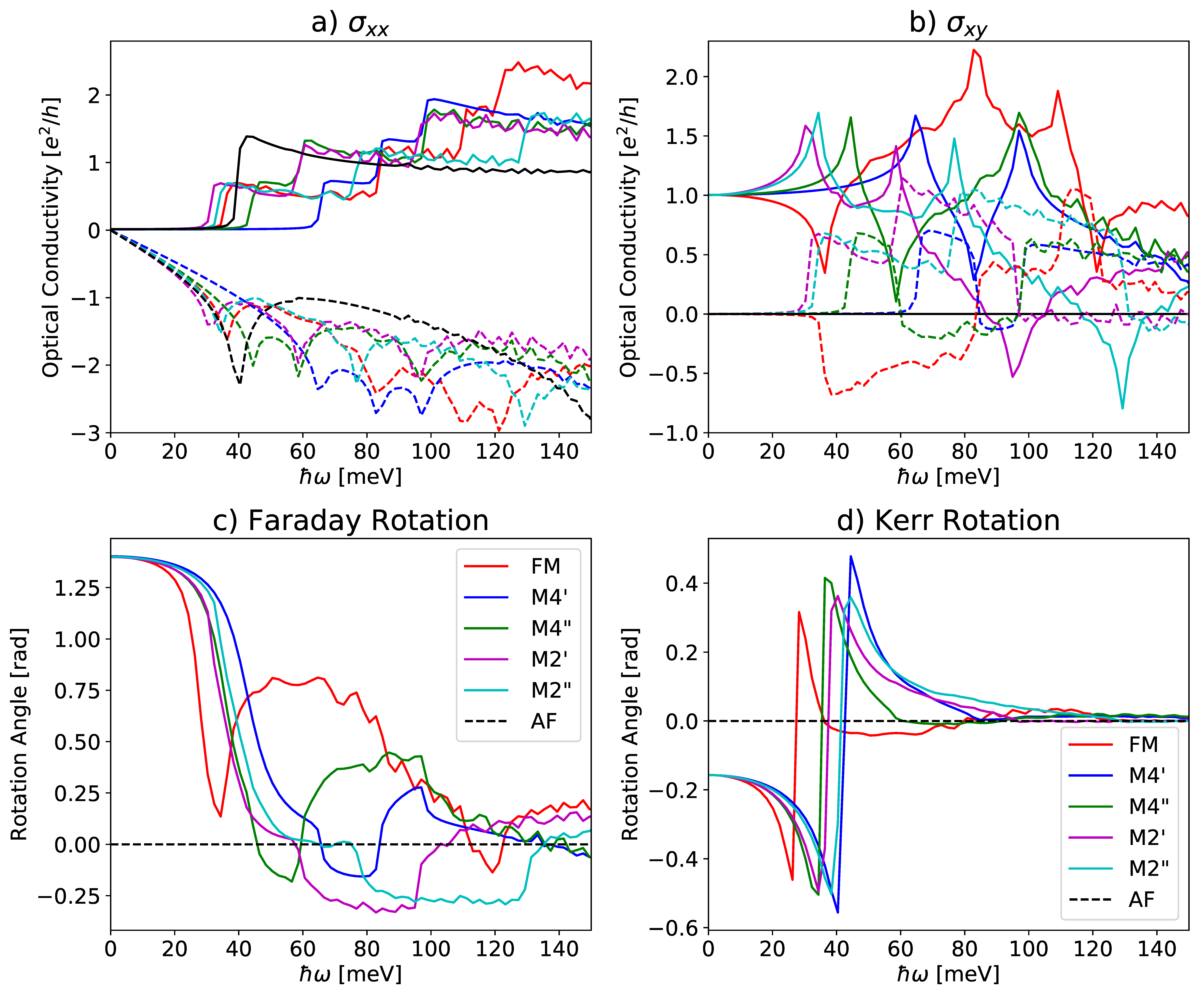}
\caption{\footnotesize Non-reciprocal optical response in 6-layer MBT thin films.
Panels (a) and (b) show plots of optical longitudinal ($\sigma_{xx}$) and  
Hall ($\sigma_{xy}$) conductivities {\it vs.} optical frequency calculated from the Kubo-Greenwood formula
using the simplified Dirac cone electronic structure model. 
In these plots the solid curves show the real part of a conductivity tensor element
while the dashed curves show the imaginary part.  Different colors 
show results for different metamagnetic states. 
The correspondence between color and metamagnetic state is repeated in 
(c) and (d), in which the Faraday and Kerr rotation angles
 in N=6 thin films are plotted {\it vs.} optical frequency.}
\label{fig:optical_response}
\end{figure}

$N=6$ layer thin films host a richer variety of metamagnetic states.
We find that all magnetic configurations with a non-zero net spin magnetization 
(FM, M2$'$/M2$''$ or M4$'$/M4$''$) have total Chern number equal to 1 and are therefore 
QAH insulators.  The states cannot be distinguished by performing DC Hall effect measurements.  Those properties that are not quantized are distinct for each of these 
states however.  For example, the Berry curvature has a different dependence on 
momentum in each case, although the total Chern number is always equal to one.
As shown in Fig. \ref{fig:optical_response}, their optical conductivities differ at finite frequencies. In Fig. \ref{fig:optical_response} (a) and (b) the real and imaginary part of longitudinal optical conductivities $\sigma_{xx} (\omega)$ and  transvere optical conductivities $\sigma_{xy} (\omega)$, calculated using the Kubo-Greenwood formula, \cite{Kubo57,Greenwood58} are shown. In these plots solid curves represent the real part of the conductivity ($\Re \sigma_{xx/xy}$) while dashed curves represent the corresponding
imaginary parts ($\Im \sigma_{xx/xy}$).  Different colors are use to 
represent different metamagnetic states. The same colors are used
for the implied frequency-dependent Faraday and Kerr rotation angles in Fig.~\ref{fig:optical_response} (c) and (d).
It follows that external magnetic fields drives the 4-layer and 6-layer thin films
from their axion insulator states to M2, M2$'$ or M2$''$ Chern insulators.

In the DC limit, all optical conductivities vanish except for $\Re \sigma_{xy}$
in the case of a QAH insulator.  $\Re \sigma_{xx}$ and $\Im \sigma_{xy}$ 
have peaks when the optical frequencies exceeds the two-dimensional
band gaps of the thin film.  $\Im \sigma_{xx}$ and $\Re \sigma_{xy}$ are, on the other hand,
non-zero for frequencies in the 
thin-film gaps.  The frequency dependence of $\Re \sigma_{xy}$ 
and $\Im \sigma_{xy}$ in the FM states differs from that of
other magnetic states in that $\Re \sigma_{xy}$ initially
decreases with frequency and $\Im \sigma_{xy}$ is negative below the band gap. 
This abnormal behavior is caused by the negative Berry curvature around the 
$\Gamma$ point in the 2D-band structure.

The optical conductivity tensor components can be converted to 
frequency-dependent Faraday and Kerr rotation angles commonly measured in experiment.
The Faraday and Kerr rotation angles are the relative rotations of left-handed and 
right-handed circularly polarized light\cite{Tse11} for transmission and 
reflection respectively, and these can be connected
to the optical conductivity by combining electromagnetic wave boundary conditions and 
Maxwell equations. The Faraday and Kerr rotation angle {\it vs.} optical frequency for 
various metamagnetic states of 6-layer thin film are shown in 
Fig.~\ref{fig:optical_response}(c) and (d) correspondingly, from which we see that 
the optical responses of all metamagnetic states with the same Chern number 
are indistinguishable in the DC limit. 
As frequency increases, the Faraday and Kerr rotation angles of different 
metamagnetic states differ substantially.  It follows that different 
metamagnetic states can be distinguished magneto-optically. 

\section{Discussion}

In our studies, we have identified the metamagnetic states that can be 
induced in MBT thin films with up to $N=6$ septuple layers by applying 
external magnetic fields.  For $N=odd$ and larger than $3$, 
all MBT thin films states are QAH insulators.  For even--$N$,
the ground states are axion insulators in the absence of an 
external magnetic field.  Metamagnetic states that are
Chern insulators can be induced by applying external magnetic field 
provided that the single-ion magnetic anisotropy is large enough 
compared with the interlayer exchange interactions. 
Both M2 and M4 states are Chern 
insulators with QAH gaps comparable to those of the FM state,
and appear at a much smaller magnetic fields. 
These metamagnetic states are not distinguished in transport experiment 
since they all have the same Chern number as the FM state. 
However, They are distinguished by their magneto-optical Kerr and 
Faraday rotation angles.

Thicker films, especially for even-layer systems and for the thin films with high 
Chern number FM states (that is $N=9$ layers thin or thicker), 
await further exploration.  Interesting questions arise when the thickness increases: is it possible, for example,
to have a high-Chern-number state at a weaker magnetic field? For MBT thin films 
the magnetic anisotropy seems to be comparable with the interlayer exchange interaction, 
it may therefore be interesting to explore other intrinsic magnetic topological insulators that have relative larger magnetic anisotropy to induce more metamagnetic states, 
or to find tools, such as electric field, that may increase the magnetic anisotropy.

\section{Acknowledgements}
RJM and OH were supported by the Center for Advancement of Topological Semimetals, an Energy Frontier Research Center funded by the U.S. Department of Energy Office of Science, Office of Basic Energy Sciences, through the Ames Laboratory under Contract No. DE-AC02-07CH11358. CL and AHM were supported by the Army Research Office under Grant Number W911NF-16-1-0472. OH gratefully acknowledges the computing resources provided on Bebop and Blues, high-performance computing clusters operated by the Laboratory Computing Resource Center at Argonne National Laboratory.

\section{Appendix}

\bibliography{MBT}

\end{document}